\documentclass[aip,apl,reprint]{revtex4-1}
\usepackage{amssymb}
\usepackage{amsmath}
\usepackage{dcolumn}
\usepackage{bm}
\usepackage{graphicx}

\begin{document}

\title{One-step implementation of multi-qubit conditional phase gating with
nitrogen-vacancy centers coupled to a high-Q silica microsphere cavity}
\author{W. L. Yang}
\affiliation{State Key Laboratory of Magnetic Resonance and Atomic
and Molecular Physics, Wuhan Institute of Physics and Mathematics,
Chinese Academy of Sciences, Wuhan 430071, China }

\author{Z. Q. Yin}
\affiliation {State Key Laboratory of Magnetic Resonance and Atomic
and Molecular Physics, Wuhan Institute of Physics and Mathematics,
Chinese Academy of Sciences, Wuhan 430071, China }
\author{Z. Y. Xu}
\affiliation{State Key Laboratory of Magnetic Resonance and Atomic
and Molecular Physics, Wuhan Institute of Physics and Mathematics,
Chinese Academy of Sciences, Wuhan 430071, China }
\affiliation{Graduate School of the Chinese Academy of Sciences,
Bejing 100049, China }

\author{M. Feng}
\email{mangfeng@wipm.ac.cn}
 \affiliation {State Key Laboratory of
Magnetic Resonance and Atomic and Molecular Physics, Wuhan Institute
of Physics and Mathematics, Chinese Academy of Sciences, Wuhan
430071, China }

\author{J. F. Du}
\email{djf@ustc.edu.cn}
\affiliation{Hefei National Laboratory for
Physics Sciences at Microscale and Department of Modern Physics,
University of Science and Technology of China, Hefei, 230026, China}

\begin{abstract}
The diamond nitrogen-vacancy (NV) center is an excellent candidate for
quantum information processing, whereas entangling separate NV centers is
still of great experimental challenge. We propose an one-step conditional
phase flip with three NV centers coupled to a whispering-gallery mode cavity
by virtue of the Raman transition and smart qubit encoding. As decoherence
is much suppressed, our scheme could work for more qubits. The experimental
feasibility is justified.
\end{abstract}

\maketitle

As a promising building block for room-temperature quantum computing,\cite%
{room} the nitrogen-vacancy (NV) center consisting of a substitutional
nitrogen atom and an adjacent vacancy in diamond can feature near-unity
quantum efficiency, a homogeneous line width, and long electronic spin
decoherence time at room-temperature,\cite{Ken}. Readout of spin state and
single qubit gating have been achieved in optical fashion in individual NV
centers,\cite{jele1} and quantum information swapping and entanglement are
available between electronic and the nuclear spins. \cite{register}

However, scalability is the main obstacle in such a system because
entanglement of NV centers in distant diamonds has never been accomplished
experimentally. Recently, Benjamin \textit{et al}\cite{Review} suggested to
entangle different NV electron spins by detecting the emitted photons, but
met some difficulties due to the particular characteristic of the NV
centers, such as the fact that $96\%$ of the emitted photons reside in broad
photon sidebands to the resonant zero phonon line (ZPL) at 637 $nm$ even in
cryogenic situation.\cite{Review} It implies that the most photons emitted
from the NV centers could not effectively interfere in the beam splitter.

We study a potential idea to entangle separate NV centers using the
quantized whispering-gallery mode (WGM) of a fused-silica high-$Q$
microsphere cavity. So far there has been much development in WGM
cavities with, such as the microtoroidal,\cite{tor}
microcylinders,\cite{lin} microdisks,\cite{disk} and
microspheres.\cite{W1} Especially, microsphere cavity had gained
widespread attention because of their ultrahigh $Q$ factor $(\geq
10^{8}$ even up to 10$^{10})$,\cite{Ar} very small volume
($V_{m}\leq 100$ $\mu m^{3}$) \cite{Ar} and simple fabrication
technique. In the fused-silica microsphere cavity, WGMs form via
total internal reflection along the curved boundary, and the small
radius of 10 $\mu m$ could lead to a vacuum electric field of 150
$V/cm$ at the sphere surface (with wavelength $600$ $nm$) and to the
$Q$ factor exceeding 10$^{9}$. On the other hand, the lowest-order
WGM corresponding to the light traveling around the equator of the
microsphere\cite{wgm} offers predominant conditions for reaching
strong coupling regime. Recent experimental progresses about the
nanocrystal-microsphere system also provide experimental evidence
for strong coupling between NV centers and the WGM of silica
microsphere \cite{park}\ or polystyrene microsphere,\cite{sch}
respectively.

The key point of our proposal is a conditional phase flip (CPF) on the NV
center electron-spins, based on recent experimental and theoretical
progresses, e.g., the possible $\Lambda$-type configuration of the optical
transitions in NV center system \cite{lambda} and the considerable
enhancement of the ZPL by embedding NV centers in some cavities.\cite%
{cavity1} We will focus on the one-step implementation of CPF gating on
three separate NV centers in diamond nanocrystals coupled to the WGM of
microsphere cavity. By virtue of the Raman transition and smartly encoding
the qubits in different NV centers, we show the effective suppression of
decoherence and the scalability of our idea.

\begin{figure}[tbph]
\centering\includegraphics[width=7 cm]{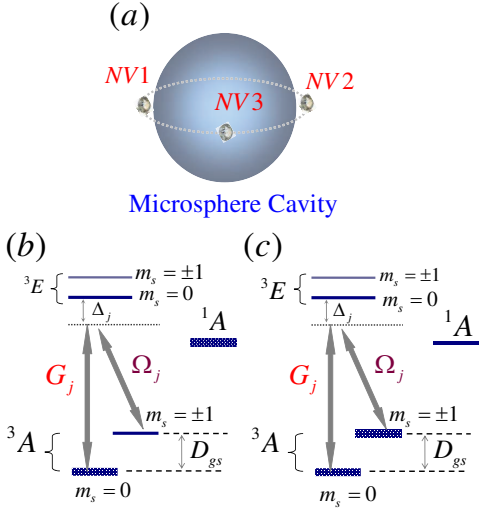} \caption{(Color
online) (a) The schematic of WGM microsphere system, where three
identical NV centers in diamond nanocrystals are equidistantly
attached around the equator of a single fused-silica microsphere
cavity. (b) The qubit definition in the first and second NV centers,
where the levels in bold encode the qubits, i.e., $\left\vert
^{3}A,m_{s}=0\right\rangle =\left\vert 1\right\rangle $ and
$\left\vert ^{1}A\right\rangle =\left\vert
0\right\rangle $. (c) The qubit definition in the third NV center, where $%
\left\vert ^{3}A,m_{s}=-1\right\rangle =\left\vert 0\right\rangle $ and $%
\left\vert ^{3}A,m_{s}=0\right\rangle =\left\vert 1\right\rangle $.}
\end{figure}

Consider three separate NV centers with each NV center attached around the
equator of a single fused-silica microsphere cavity. Each NV center is
negatively charged with two unpaired electrons located at the vacancy,
usually treated as electron spin-$1$. So the ground state is spin triplet
and labeled as $^{3}A$, with the levels $m_{S}=\pm 1$ nearly degenerated and
a zero-field splitting ($D_{gs}=$2.87 GHz) between the states $m_{S}=0$ and $%
m_{S}=\pm 1$.\cite{manson} The excited state $^{3}E$ is also a spin triplet,
associated with a broadband photoluminescence emission with ZPL of 1.945 $eV$%
, which allows optical detection of individual NV defects using confocal
microscopy. For clarity of description, we adopt following denotations: $%
\left\vert ^{3}A,m_{s}=0\right\rangle =\left\vert g\right\rangle$, $%
\left\vert ^{3}A,m_{s} =-1\right\rangle=\left\vert e\right\rangle$, $%
\left\vert ^{3}E,m_{s}=0\right\rangle=\left\vert E\right\rangle$ and $%
\left\vert ^{1}A\right\rangle=\left\vert f\right\rangle$. By combining
individual laser pulse irradiation ($\sigma ^{+}$ circularly polarized and
coupling strength $\Omega _{j}$ \cite{lambda,opt}) with the WGM field ($%
\sigma ^{0}$ polarized and coupling strength $G_{j}$), one can model each NV
center as a $\Lambda $-type three-level structure, whose Hamiltonian under
the rotating wave approximation can be written in units of $\hbar =1$ as $%
H_{I}=\sum\nolimits_{j=1}^{3}[\Delta _{j}\left\vert E_{j}\right\rangle
\left\langle E_{j}\right\vert +(G_{j}a^{+}\left\vert g_{j}\right\rangle
\left\langle E_{j}\right\vert +\Omega _{j}\left\vert E_{j}\right\rangle
\left\langle e_{j}\right\vert +H.c.)]$, where $a^{+}(a)$ is the creation
(annihilation) operator for the WGM. Applying standard quantum optical
techniques,\cite{Gar} under the large-detuning conditions $\left\vert \Delta
_{j}\right\vert \gg $ $\left\vert \Omega _{j}\right\vert $, $\left\vert
G_{j}\right\vert $, we could adiabatically eliminate the virtually excited
states $\left\vert E_{j}\right\rangle$, which yields the effective
Hamiltonian \cite{ex1,ex2,Lu} $H_{eff}^{^{\prime }}=\sum\nolimits_{j=1}^{3}[
\frac{\Omega _{j}^{2}}{\Delta _{j}}\left\vert e_{j}\right\rangle
\left\langle e_{j}\right\vert +\frac{G_{j}^{2}}{\Delta _{j}}a^{+}a\left\vert
g_{j}\right\rangle \left\langle g_{j}\right\vert +\frac{G_{j}\Omega _{j}}{%
\Delta _{j}}(\left\vert e_{j}\right\rangle \left\langle g_{j}\right\vert
a+\left\vert g_{j}\right\rangle \left\langle e_{j}\right\vert a^{+})]$,
where $G_{j}\Omega _{j}/\Delta _{j}$ is the effective Rabi frequency. The
first two terms in $H_{eff}^{^{\prime }}$ represent laser-induced and
photon-induced dynamic energy shifts, respectively, among which the
photon-induced level shifts can be eliminated when the WGM cavity is
initially prepared in the vacuum state $\left\vert 0_{c}\right\rangle $, and
the laser-induced level shifts can be compensated straightforwardly using
additional lasers with appropriate frequencies.\cite{ex1,Lu} So the
effective Hamiltonian can be further reduced to $H_{eff}=\sum%
\nolimits_{j=1}^{3}\tilde{G}_{j}[\left\vert e_{j}\right\rangle \left\langle
g_{j}\right\vert a+\left\vert g_{j}\right\rangle \left\langle
e_{j}\right\vert a^{+}],$ with $\tilde{G}_{j}=G_{j}\Omega _{j}/\Delta _{j}$.

The three-qubit CPF gate $U_{CPF}=diag\{1,1,1,1,1,1,1,-1\}$ is carried out
in the computational subspace spanned by \{$\left\vert
g_{1}g_{2}g_{3}\right\rangle ,$ $\left\vert g_{1}g_{2}e_{3}\right\rangle ,$ $%
\left\vert g_{1}f_{2}g_{3}\right\rangle ,$ $\left\vert
g_{1}f_{2}e_{3}\right\rangle ,$ $\left\vert f_{1}g_{2}g_{3}\right\rangle ,$ $%
\left\vert f_{1}g_{2}e_{3}\right\rangle ,$ $\left\vert
f_{1}f_{2}g_{3}\right\rangle ,$ $\left\vert f_{1}f_{2}e_{3}\right\rangle $%
\}, where the qubit definition is sketched in Fig. 1. Note that the
effective resonant interactions only occur between the states $\left\vert
g_{j}\right\rangle $ and $\left\vert e_{j}\right\rangle $. So the auxiliary
states $\left\vert f_{j}\right\rangle $ are not involved in the interaction
throughout our scheme. As a result, the states $\left\vert
g_{1}g_{2}g_{3}\right\rangle,$ $\left\vert g_{1}f_{2}g_{3}\right\rangle,$ $%
\left\vert f_{1}g_{2}g_{3}\right\rangle$, and $\left\vert
f_{1}f_{2}g_{3}\right\rangle$ remain unchanged in the evolution.

We first assume the system to be initially in the state $\left\vert
f_{1}f_{2}e_{3}\right\rangle \left\vert 0_{c}\right\rangle $ with $%
\left\vert 0_{c}\right\rangle $ $(\left\vert 1_{c}\right\rangle)$ the vacuum
(one-photon) state of the WGM field. Then only the $3$rd NV center evolves
under $H_{eff}$, that is, $\left\vert f_{1}f_{2}e_{3}\right\rangle
\left\vert 0_{c}\right\rangle \longrightarrow [\cos (\tilde{G}%
_{3}t)\left\vert f_{1}f_{2}e_{3}\right\rangle \left\vert 0_{c}\right\rangle
-i\sin (\tilde{G}_{3}t)\left\vert f_{1}f_{2}g_{3}\right\rangle \left\vert
1_{c}\right\rangle ]$.

Next, we consider another situation with the initial state $\left\vert
g_{k}f_{j}e_{3}\right\rangle \left\vert 0_{c}\right\rangle $ $(k,j=$1, 2$%
,k\neq j)$, for which the $j$th qubit remains unchanged in the interaction.
So we have $\left\vert g_{k}f_{j}e_{3}\right\rangle \left\vert
0_{c}\right\rangle \longrightarrow \tilde{N}_{1}\{[\tilde{G}_{3}^{2}\cos (%
\tilde{G}_{k}^{^{\prime }}t)+\tilde{G}_{k}^{2}] \times \left\vert
g_{k}f_{j}e_{3}\right\rangle \left\vert 0_{c}\right\rangle +\tilde{G}_{3}%
\tilde{G}_{k}[\cos (\tilde{G}_{k}^{^{\prime }}t)-1]\left\vert
e_{k}f_{j}g_{3}\right\rangle \left\vert 0_{c}\right\rangle -i\tilde{G}_{3}%
\tilde{G}_{k}^{^{\prime }}\sin (\tilde{G}_{k}^{^{\prime }}t)\left\vert
g_{k}f_{j}g_{3}\right\rangle \left\vert 1_{c}\right\rangle \}$, where $%
\tilde{N}_{1}=1/\tilde{G}_{k}^{^{\prime }2}$ with $\tilde{G} _{k}^{^{\prime
}}=\sqrt{\tilde{G}_{k}^{2}+\tilde{G}_{3}^{2}}.$

To achieve our aim, we need further consider an initial state $\left\vert
g_{1}g_{2}e_{3}\right\rangle \left\vert 0_{c}\right\rangle $, which evolves
as $\left\vert g_{1}g_{2}e_{3}\right\rangle \left\vert 0_{c}\right\rangle
\longrightarrow \tilde{N}_{2}\{[\tilde{G}_{3}^{2}\cos (\tilde{G}^{^{\prime
\prime }}t)+\tilde{G}_{1}^{2}+\tilde{G}_{2}^{2}] \times \left\vert
g_{1}g_{2}e_{3}\right\rangle \left\vert 0_{c}\right\rangle +\tilde{G}%
_{3}[\cos (\tilde{G}^{^{\prime \prime }}t)-1](\tilde{G}_{1}\left\vert
e_{1}g_{2}g_{3}\right\rangle \left\vert 0_{c}\right\rangle +\tilde{G}%
_{2}\left\vert g_{1}e_{2}g_{3}\right\rangle \left\vert 0_{c}\right\rangle )-i%
\tilde{G}_{3}\tilde{G}^{^{\prime \prime }}\sin (\tilde{G}^{^{\prime \prime
}}t)\left\vert g_{1}g_{2}g_{3}\right\rangle \left\vert 1_{c}\right\rangle \}$%
, where $\tilde{N}_{2}=1/\tilde{G}^{^{\prime \prime }2}$ with $\tilde{G}
^{^{\prime \prime }}=\sqrt{\sum\nolimits_{j=1}^{3}\tilde{G}_{j}^{2}}$.

We assume that coupling strengths satisfy the condition $\tilde{G}_{1}=
\tilde{G}_{2}\gg \tilde{G}_{3},$ and the interaction time is $%
t_{0}=(2k+1)\pi /\tilde{G}_{3}$ with $k$ the non-negative integers. Then
above equations for state evolution can be greatly simplified to $\left\vert
f_{1}f_{2}e_{3}\right\rangle \left\vert 0_{c}\right\rangle \longrightarrow
-\left\vert f_{1}f_{2}e_{3}\right\rangle \left\vert 0_{c}\right\rangle $, $%
\left\vert g_{k}f_{j}e_{3}\right\rangle \left\vert 0_{c}\right\rangle
\longrightarrow \beta \left\vert g_{k}f_{j}e_{3}\right\rangle \left\vert
0_{c}\right\rangle,$ $\left\vert g_{1}g_{2}e_{3}\right\rangle \left\vert
0_{c}\right\rangle \longrightarrow \alpha \left\vert
g_{1}g_{2}e_{3}\right\rangle \left\vert 0_{c}\right\rangle$, where $\alpha
=[m^{2}\cos (\sqrt{m^{2}+2}\pi /m)+2]/(m^{2}+2)$ and $\beta =[m^{2}\cos (%
\sqrt{m^{2}+1}\pi /m)+1]/(m^{2}+1)$ and $m=\tilde{G}_{3}/\tilde{G}_{i}$ with
$\allowbreak i=$1, 2. In the case of $m\ll 1$, we have $\beta \approx \alpha
\approx 1$ and thereby obtain an nearly perfect three-qubit CPF gate $%
U_{CPF}^{^{\prime }}=diag\{1,\alpha ,1,\beta ,1,\beta ,1,-1\}$, where $%
\alpha =0.999247$ and $\beta =0.999879$ in the case of $m=0.1$.

\begin{figure}[tbph]
\centering\includegraphics[width=8.4 cm]{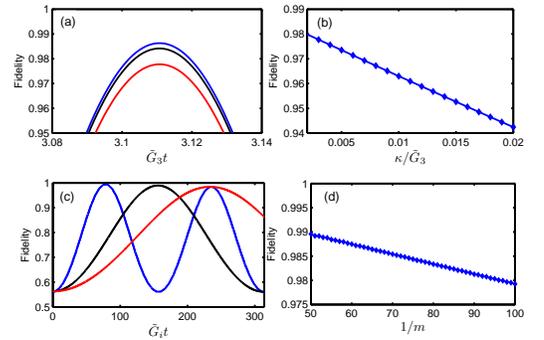} \caption{(Color
online) The fidelity of the three-qubit CPF gate versus
different parameters in the case of $\Gamma_{eg}=\tilde{G}_{3}/100$ and $%
\Gamma_{fg}=\tilde{G}_{3}/10^{6}$: (a) versus $\tilde{G}_{3}t$ where $m=0.1$%
, the blue, black, red lines denote the case of $\protect\kappa =\tilde{G}%
_{3}/100,$ $\tilde{G}_{3}/50,$ and $\tilde{G}_{3}/20,$ respectively;
(b)
versus $\protect\kappa /\tilde{G}_{3}$ where $m=0.1$; (c) versus $\tilde{G}%
_{i}t$ where $i=$1, 2, $\protect\kappa =\tilde{G}_{3}/100$, the
blue, black, red lines denote the cases of $m=1/25$, $1/50$, and
$1/75,$ respectively; (d) versus $m$ where $\protect\kappa
=\tilde{G} _{3}/100$.}
\end{figure}

In practice, decoherence places limits in the process discussed above. In
our scheme, the decoherence rate includes the radiative decay $\Gamma_{eg}$
between $\left\vert e\right\rangle$ and $\left\vert g\right\rangle$, the
radiative decay $\Gamma_{fg}$ between $\left\vert f\right\rangle $ and $%
\left\vert g\right\rangle $, WGM field decay rate $\kappa$ and the decay
from $|E \rangle$. Due to the large detuning, however, we may put the decay
from $|E \rangle$ aside, but model the effect of other decoherence by the
Lindblad equation,
\begin{eqnarray}
\dot{\rho} &=&-i[H,\rho ]+\kappa (2a\rho a^{+}-a^{+}a\rho -\rho a^{+}a)
\notag \\
&&+\sum\nolimits_{j=1}^{3}\{\Gamma _{eg}(2\sigma _{ge}^{j}\rho \sigma
_{eg}^{j}-\sigma _{eg}^{j}\sigma _{ge}^{j}\rho -\rho \sigma _{eg}^{j}\sigma
_{ge}^{j})  \notag \\
&&+\Gamma _{fg}(2\sigma _{gf}^{j}\rho \sigma _{fg}^{j}-\sigma
_{fg}^{j}\sigma _{gf}^{j}\rho -\rho \sigma _{fg}^{j}\sigma _{gf}^{j})\},
\end{eqnarray}
with $\sigma _{ge}^{j}=\left\vert g_{j}\right\rangle \left\langle
e_{j}\right\vert $ and $\sigma _{gf}^{j}=\left\vert g_{j}\right\rangle
\left\langle f_{j}\right\vert $. In this composite nanocrystal-microsphere
system, the coupling strength between NV center and WGM could be \cite{WG} $%
G_{\max }=\Gamma _{0}\sqrt{V_{a}/V_{m}}/2$, where $V_{m}$ is the
cavity-electromagnetic-mode volume, and $V_{a}=3c\lambda ^{2}/4\pi \Gamma
_{0}$ denotes a characteristic interaction volume with $\lambda $ the
transition wavelength between states $\left\vert E\right\rangle $ and $%
\left\vert g\right\rangle $, $\Gamma _{0}$ the spontaneous decay rate of the
excited state $\left\vert E\right\rangle$ and $c$ the speed of light. Using
the values $\lambda =$637 $nm$, $\Gamma _{0}=$2 $\pi \times 83$ MHz,\cite{NJ}
and $V_{m}=$20 $\mu m^{3}$, we have the maximal coupling strength $G_{\max
}\simeq $2$\pi \times 5.5$ GHz. In the case of $G_{_{1}}=G_{_{2}}=G_{3}=G_{%
\max}$, $\Omega _{_{1}}=\Omega _{_{2}}=\Omega _{\max }=$2$\pi \times 2.5$
GHz, $\Omega _{_{3}}=\Omega _{\max }/10=$2$\pi \times 250$ MHz, and the
detuning $\Delta _{j}=$2$\pi \times 25$ GHz, the effective coupling rates
are $\tilde{G}_{1}=$ $\tilde{G}_{2}\simeq$2$\pi \times 550$ MHz, and $\tilde{%
G}_{3}\simeq $2$\pi \times 55$ MHz. Moreover, the WGM field decay rate,
given the cavity quality factor $Q=10^{9}$ and $\kappa =c/\lambda Q=$2$\pi
\times 0.5$ MHz, is about $\tilde{G}_{3}/100$, which implies a pretty small
detrimental effect on our scheme. In addition, the characteristic
spontaneous decay rate $\Gamma _{eg}$ regarding the excited state to $%
\left\vert e\right\rangle $ and $\left\vert g\right\rangle $ could be
estimated as $\Gamma _{0}\Omega _{j}g_{j}/\Delta _{j}^{2}\simeq $2$\pi
\times 0.83$ MHz.\cite{Spo,ex} Fig. 2 plots the fidelity of the three-qubit
CPF gate $U_{CPF}^{^{\prime}}$ in the decay case. With $(\kappa$, $%
\Gamma_{eg}$, $\Gamma_{fg})\ll \tilde{g}_{3}$, high fidelity of the $%
U_{CPF}^{^{\prime }}$ can be substantially retained.

In realistic experiments, the electron-spin relaxation time $T_{1}$ of the
diamond NV center ranges from $6$ ms at room temperature \cite{mani1} to
seconds at low temperature. In addition, the dephasing time $T_{2}=$350 $\mu$%
s induced by the nuclear spin fluctuation inside the NV center has been
reported.\cite{ini1} Moreover, $|E \rangle$ experiences spin-orbit coupling
and thereby the electron-phonon coupling, affecting the orbit, leads to
dephasing of $|E \rangle$, particularly in room temperature situation.
Fortunately, due to large detuning, there are only 2.2$\%$$%
(=G_{max}\Omega_{max}/\Delta^{2})$ probability with $|E \rangle$ to be
populated, implying an effective dephasing time $1/(\Gamma_{0}\times
2.2\%)=0.87 \mu$s. In contrast, our three-qubit CPF gating time is $%
t_{0}=\pi /\tilde{G}_{3}=$0.009 $\mu $s. As a result, nearly 100 gate
operations are feasible under present experimental conditions. This also
implies that the influence from the intrinsic damping and dephasing is
negligible in our scheme.

In conclusion, we have investigated an one-step conditional gate with three
NV centers coupled to a WGM cavity by virtue of the Raman transition and by
smart qubit encoding. Straightforward extension of the idea to more NV
centers is possible if the NV centers are appropriately coupled to the WGM
of microsphere cavity.

The work is supported by the NNSFC, CAS and NFRPC.

\newpage

\end{document}